\documentclass{hep99_preprint}

\usepackage{epsfig}

\def\d0{D\O}
\def\D0{D\O}

\def\wz{$WZ$}
\def\wwg{$WW\gamma$}

\def\wwz{$WWZ$}

\def\etmisv {\mbox{${\hbox{${\vec E}$\kern-0.6em\lower-.1ex\hbox{/}}}_T$}}
\def\etmis  {\mbox{${\hbox{$E$\kern-0.6em\lower-.1ex\hbox{/}}}_T$}}

\def\ifmath#1{\relax\ifmmode #1\else $#1$}%
\def\TeV{\ifmmode {\mathrm{ Te\kern -0.1em V}}\else
                   \textrm{Te\kern -0.1em V}\fi}%
\def\GeV{\ifmmode {\mathrm{ Ge\kern -0.1em V}}\else
                   \textrm{Ge\kern -0.1em V}\fi}%
\def\MeV{\ifmmode {\mathrm{ Me\kern -0.1em V}}\else
                   \textrm{Me\kern -0.1em V}\fi}%
\def\GeVcc{\ifmmode {\mathrm{ \GeV/c^2}}\else
                   \textrm{Ge\kern -0.1em V/c$^2$}\fi}%
\def\MeVcc{\ifmmode {\mathrm{ \MeV/c^2}}\else
                   \textrm{Me\kern -0.1em V/c$^2$}\fi}%

\def\Aslash{\mbox{${\hbox{$A$\kern-0.55em\hbox{/}}}$}}
\def\pslash{\mbox{${\hbox{$p$\kern-0.45em\hbox{/}}}$}}

\def\to{\rightarrow}
\def\S{\mathhexbox279}
\def\gesim{\,{\raise-3pt\hbox{$\sim$}}\!\!\!\!\!{\raise2pt\hbox{$>$}}\,}
\def\lesim{\,{\raise-3pt\hbox{$\sim$}}\!\!\!\!\!{\raise2pt\hbox{$<$}}\,}
\def\boldoverdot{\,{\raise6pt\hbox{\bf.}\!\!\!\!\>}}

\def\pcal{{\cal P}}

\def\diag{\hbox{\diag}}

\def\doubleundertext#1{
{\undertext{\vphantom{y}#1}}\par\nobreak\vskip-\the\baselineskip\vskip4pt%
\undertext{\hbox to 2in{}}}
\def\inbox#1{\vbox{\hrule\hbox{\vrule\kern5pt
     \vbox{\kern5pt#1\kern5pt}\kern5pt\vrule}\hrule}}
\def\sqr#1#2{{\vcenter{\hrule height.#2pt
      \hbox{\vrule width.#2pt height#1pt \kern#1pt
         \vrule width.#2pt}
      \hrule height.#2pt}}}

\def\today{\ifcase\month\or
  January\or February\or March\or April\or May\or June\or
  July\or August\or September\or October\or November\or December\fi
  \space\number\day, \number\year}
\def\pmb#1{\setbox0=\hbox{#1}%
  \kern-.025em\copy0\kern-\wd0
  \kern.05em\copy0\kern-\wd0
  \kern-.025em\raise.0433em\box0 }
\def\sumprime_#1{\setbox0=\hbox{$\scriptstyle{#1}$}
  \setbox2=\hbox{$\displaystyle{\sum}$}
  \setbox4=\hbox{${}'\mathsurround=0pt$}
  \dimen0=.5\wd0 \advance\dimen0 by-.5\wd2
  \ifdim\dimen0>0pt
  \ifdim\dimen0>\wd4 \kern\wd4 \else\kern\dimen0\fi\fi
\mathop{{\sum}'}_{\kern-\wd4 #1}}
\begin{document} 

\title{\boldmath $W$ and $Z$ Properties at the Tevatron$^\dag$}

\author{John~Ellison\\
{\it \small (for the CDF and \d0\ Collaborations)}}
\address{Department of Physics, University of California, Riverside, CA 92521\\[3pt]
E-mail: john.ellison@ucr.edu}

\abstract{We present recent results from CDF and \d0\ on $W$ and $Z$
production cross sections, the width of the $W$ boson, 
$\tau - e$ universality in $W$ decays, trilinear gauge boson couplings, and
on the observation of $Z \to b \bar b$. }

\maketitle

\fntext{\dag}{Paper presented at the International Europhysics Conference on High Energy Physics, Tampere, Finland, 15-21 July, 1999.}

\section{Introduction}

In this paper we review some recent results on $W$ and $Z$ properties
obtained by the CDF and \d0\ collaborations at the Fermilab
Tevatron. The results are based on data sets collected during the 1994-95
run (``Run 1b''), with total integrated luminosities of $\approx
85-90$~pb$^{-1}$ per experiment. CDF (\d0) observed 41,666 (67,078)
$W \to e \nu$ candidate events and 5,152 (5,397) $Z \to e^+ e^-$
candidates.

\section{\boldmath Measurement of the Ratio of $W \to e \nu$ and $Z \to e^+ e^-$ 
Cross Sections}

\begin{figure}[h]
\vspace{-.5cm}
 \epsfysize = 7.5cm
    \centerline{\epsffile{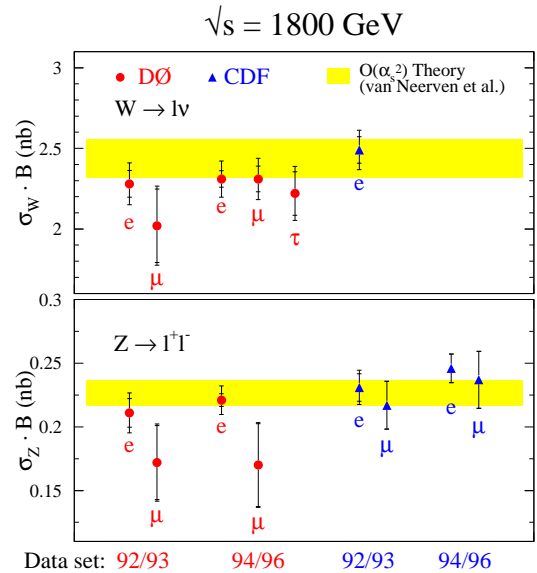}}
\caption{Measurements of the $W \to e \nu$ and $Z \to e^+ e^-$ cross 
sections from \d0\ and CDF.}
\label{fig:xsec}
\end{figure}

New results on the $W$ and $Z$ production cross sections times
electronic branching ratios from CDF and \D0\ are shown in
Fig.~\ref{fig:xsec}.  \D0\ measure~\cite{d0_wzxs} $\sigma_W \cdot B(W \to e \nu) =
2310 \pm 10~\mathrm{(stat)} \pm 50~\mathrm{(syst)} \pm
100~\mathrm{(lum)}$~pb and $\sigma_Z \cdot B(Z \to e^+ e^-) = 221 \pm
3~\mathrm{(stat)} \pm 4~\mathrm{(syst)} \pm 10~\mathrm{(lum)}$~pb, where
``lum'' is due to the uncertainty on the integrated luminosity. CDF
obtain $\sigma_Z \cdot B(Z \to e^+ e^-) = 249 \pm 5~\mathrm{(stat \oplus syst)} \pm
10~\mathrm{(lum)}$~pb 
and $\sigma_Z \cdot B(Z \to \mu^+ \mu^-) = 237 \pm 9~\mathrm{(stat \oplus lum)} \pm
9~\mathrm{(lum)}$~pb~\cite{cdf_zxs}. 
%
%

The errors are dominated by the uncertainty in the integrated
luminosity of the data samples. The \d0\ and CDF results must be
compared with care, since the experiments use different total $p \bar
p$ cross sections to determine their integrated luminosities. CDF use
their own measurement, while \d0\ take the average of the CDF, E710 and
E811 measurements. As a result there is a scale factor which must be
applied to the measured cross sections: e.g. if using the CDF
normalization, the \d0\ Run 1b cross sections must be multiplied by
1.062. Note that the results in Fig.~\ref{fig:xsec}
have {\it not} been rescaled.

The integrated luminosity uncertainty and many of the other systematic
errors cancel in the ratio of cross sections $R = \sigma_W \cdot B(W
\to e \nu) / \sigma_Z \cdot B(Z \to e^+ e^-)$. This allows indirect,
precise measurements of the $W \to e \nu$ branching fraction and the
width of the $W$ boson. This follows using
\begin{eqnarray*}
{\sigma_W \cdot B(W\to e \nu) \over {\sigma_Z \cdot B(Z \to e^+ e^-)}} =
{\sigma_W \over \sigma_Z} {1 \over {B(Z \to e^+ e^-)}}
{\Gamma (W \to e \nu) \over {\Gamma(W)}}
\end{eqnarray*}
\noindent
together with the theoretical calculation of $\sigma_W / \sigma_Z$~\cite{vanNeerven},
the measured $Z \to e^+ e^-$ branching ratio from
LEP~\cite{pdg}, and the SM value of $\Gamma (W \to e \nu)$~\cite{Rosner}.

\begin{figure}
 \epsfysize = 7cm
    \centerline{\epsffile{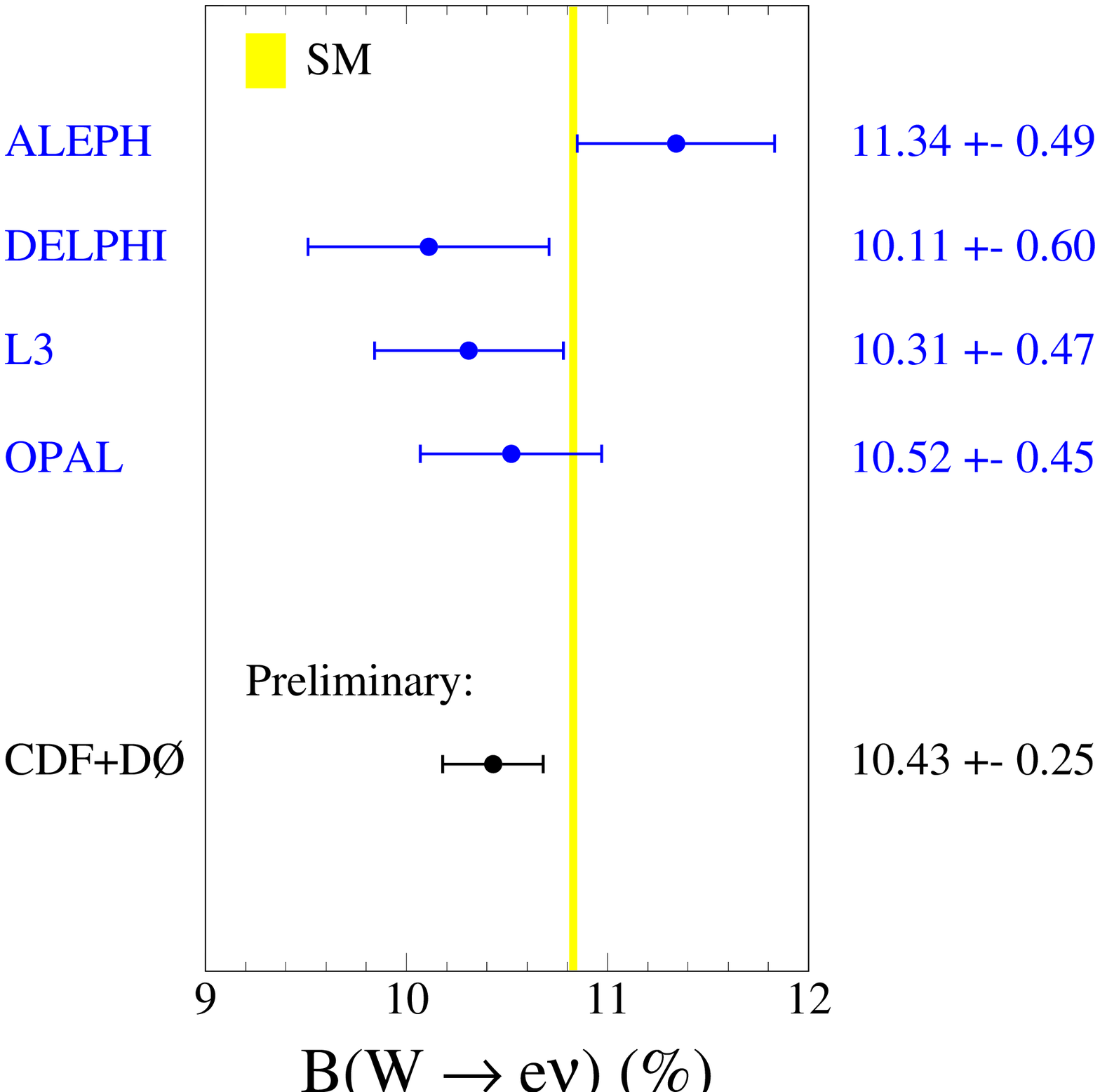}}
\caption{Measurements of $B(W \to e \nu)$.}
\label{fig:wbr}
\end{figure}

\begin{figure}
 \epsfysize = 7cm
    \centerline{\epsffile{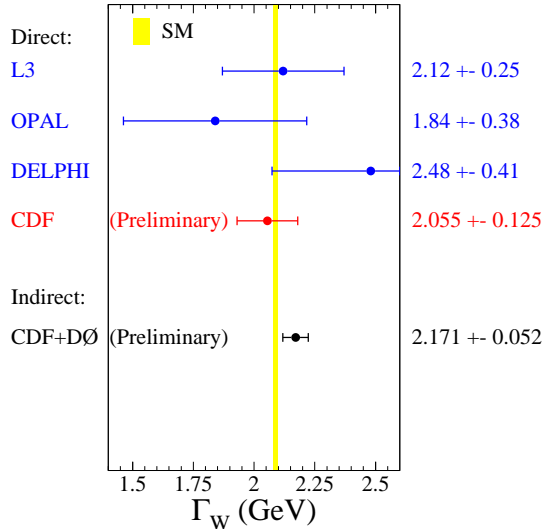}}

\caption{Direct and indirect measurements of the $W$ boson width.}
\label{fig:wwidth}
\end{figure}

The measured values of $R$ are $R = 10.49 \pm 0.14~\mathrm{(stat)} \pm
0.21~\mathrm{(syst)}$ for \d0\ and $R = 10.38 \pm 0.14~\mathrm{(stat)}
\pm 0.17~\mathrm{(syst)}$ for CDF, using the combined electron data
from Runs 1a and 1b. The main sources of systematic errors are due to
uncertainties in backgrounds, efficiencies, and electron energy scale.
A 1\% error due to NLO electroweak radiative corrections is also
included.  The two $R$ measurements have been combined, yielding $R =
10.42 \pm 0.18$. Using this combined value of $R$, the resulting
branching fraction is $B(W \to e \nu) = (10.43 \pm 0.25)$\% and the
width of the $W$ boson is determined to be $\Gamma(W) =
2.171 \pm 0.052$~GeV.  The results agree with the SM predictions when
the errors are taken into account, as shown in Figs.~\ref{fig:wbr} and
\ref{fig:wwidth}.  A significant source of systematic error (1.5\%)
arises from the theoretical uncertainty on $\sigma_W / \sigma_Z$ due
to the choice of renormalization scheme and electroweak radiative
corrections.  Note that at present the errors due to theoretical
uncertainties on $B(W \to e \nu)$ and $\Gamma(W)$ are larger than the
statistical uncertainty.

\begin{figure}
 \epsfysize = 6.5cm
    \centerline{\epsffile{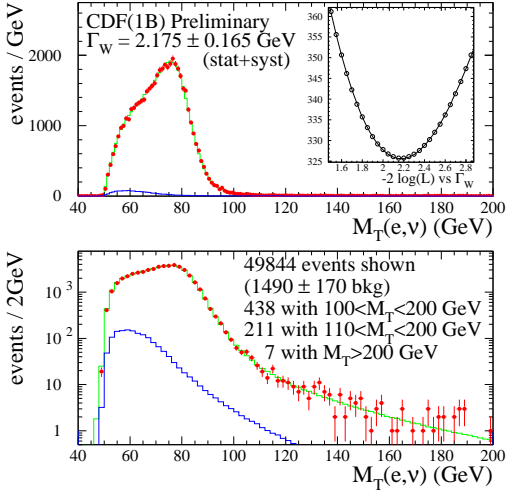}}
\caption{CDF direct measurement of the $W$ boson width using a fit to 
the $e \nu$ transverse mass.}
\label{fig:wdirect_e}
\end{figure}

\begin{figure}
 \epsfysize = 6.5cm
    \centerline{\epsffile{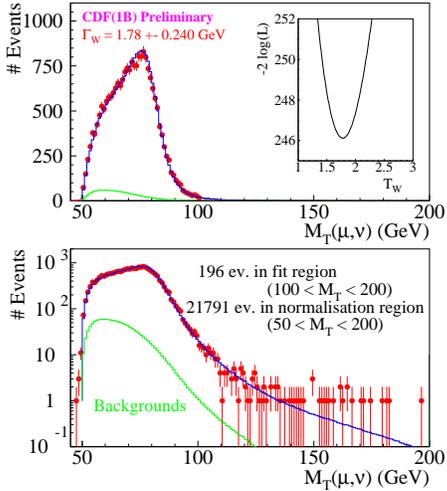}}
\caption{CDF direct measurement of the $W$ width using a fit to 
the $\mu \nu$ transverse mass.}
\label{fig:wdirect_mu}
\end{figure}

A direct measurement of the $W$ boson width is possible using a fit to
transverse mass ($M_T$) spectrum in $W$ events. The $W$ width directly affects
the shape of the distribution, most prominently at high values of
$M_T$, where the Breit-Wigner line shape dominates over detector
resolution effects. CDF have new preliminary results for Run 1b $W \to
e \nu$ and $W \to \mu \nu$ events, using a binned likelihood fit in
the region $M_T > 100$~GeV$/c^2$. The $W$ events are modeled using a
similar simulation to that used in the $W$ mass analyses~\cite{Carithers}.
This method is less model-dependent than the
indirect measurement discussed above, but with the current data sets
it is statistically limited. The transverse mass fits are shown in
Figs.~\ref{fig:wdirect_e} and \ref{fig:wdirect_mu}. The results are
$\Gamma(W) = 2.17 \pm 0.125~\mathrm{(stat)} \pm
0.105~\mathrm{(syst)}$~GeV from the electron data and $\Gamma(W) =
1.78 \pm 0.195~\mathrm{(stat)} \pm 0.135~\mathrm{(syst)}$~GeV from the
muon data. These results are combined with the Run 1a electron
measurement, yielding $\Gamma(W) = 2.055 \pm 0.100~\mathrm{(stat)} \pm
0.075~\mathrm{(syst)}$~GeV. This result is consistent with the SM
prediction, as shown in Fig.~\ref{fig:wwidth}.

\section{\boldmath $\tau-e$ universality in $W$ decays}

\begin{figure}
 \epsfysize = 6.5cm
    \centerline{\epsffile{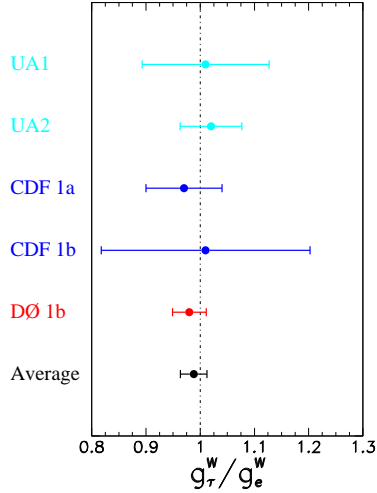}}
\caption{Measurements of $g_\tau^W / g_e^W$.}
\label{fig:gtau}
\end{figure}

\D0\ have updated their preliminary measurement of the $W$
cross section times $W \to \tau \nu$ branching ratio using Run 1b data and the new
luminosity normalization, and obtain $\sigma_W \cdot B(W \to \tau \nu) =
2220 \pm 90~\mathrm{(stat)} \pm 100~\mathrm{(syst)} \pm
100~\mathrm{(lum)}$~pb. The ratio of this quantity to the
corresponding electron-channel quantity measures the ratio of the
electroweak charged current couplings, $g_\tau^W / g_e^W$. \d0\
measure $g_\tau^W / g_e^W = 0.98 \pm 0.03$, to be compared with the
earlier CDF result of $g_\tau^W / g_e^W = 0.97 \pm 0.07$ using Run 1a
data, and the preliminary CDF result $g_\tau^W / g_e^W = 1.01 \pm
0.19$ which uses Run 1b data, but is obtained from an analysis based
on the difference in electron impact parameter distributions in $W \to
\tau \nu \to e \nu \nu \nu$ and $W \to e \nu$ events. These results
are shown in Fig.~\ref{fig:gtau}.

\section{Updated \d0\ results on trilinear gauge boson couplings}

\d0\ have recently published limits on the trilinear gauge boson
couplings from $p \bar p \to WW/WZ \to \mu \nu \, jj$ events~\cite{d0_wztrilep_prd}.
They use a likelihood fit to the $p_T (\mu
\nu)$ spectrum, which is sensitive to non-SM couplings.  Anomalous
couplings would result in an enhancement of events at high $p_T (\mu
\nu)$. No such enhancement is seen and \d0\ obtain 95\% confidence
level limits, assuming equal \wwg\ and \wwz\ couplings (i.e.
$\lambda_\gamma = \lambda_Z$ and $\Delta \kappa_\gamma = \Delta
\kappa_Z$) using a dipole form factor with form factor scale
$\Lambda_{FF} = 2$~TeV: $-0.52 < \lambda < 0.54$ and $-0.62 < \Delta
\kappa < 0.88$.

\begin{figure}
 \epsfysize = 7cm
    \centerline{\epsffile{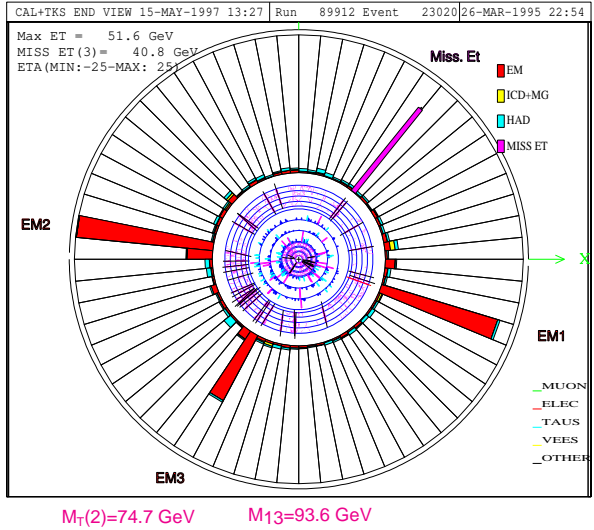}}
\caption{End view of the \d0\ calorimeter and tracking systems showing a candidate $WZ \to e \nu \, ee$ decay.}
\label{fig:wzevt}
\end{figure}

\D0\ have also searched for \wz\ events in the $e \nu \, ee$ and $\mu
\nu \, ee$ channels ~\cite{d0_wztrilep_prd}. One event passes the
selection criteria, an $e \nu \, ee$ candidate, shown in
Fig.~\ref{fig:wzevt}. For both channels combined, the SM prediction is
$0.245 \pm 0.0154$ events, with an estimated background of $0.50 \pm
0.15$ events. In the absence of an excess of events, which would be an
indication of non-SM \wwz\ couplings, \d0\ set limits on anomalous
couplings. This analysis is most sensitive to the couplings $\lambda$
and $\Delta g_1^Z$ , with resulting limits $\left| \lambda \right| <
1.42$ and $\left| \Delta g_1^Z \right| < 1.63$ at the 95\% CL, using a
form factor scale of 1~TeV. Because \wz\ production is sensitive only
to the \wwz\ coupling, these results are independent of any
assumptions about the \wwg\ coupling.

\begin{figure}
 \epsfysize = 10cm
    \centerline{\epsffile{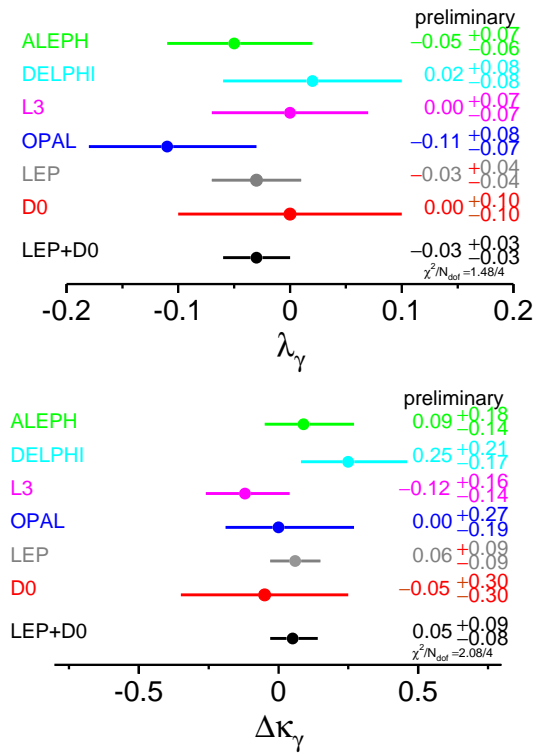}}
\caption{Measurements of the trilinear gauge boson coupling parameters $\lambda$ and 
$\Delta \kappa$ from \d0\ and LEP.}
\label{fig:tgc}
\end{figure}

The new results on the trilinear gauge boson couplings have been
combined with previous results from \d0\ to yield new global limits
from \d0\ utilizing all diboson production
analyses. Figure~\ref{fig:tgc} shows the results for the couplings
$\Delta \kappa$ and $\lambda$. The \d0\ results use a form factor
scale of 2~TeV, and therefore, the LEP errors should be scaled by a
factor of $(1 + s / \Lambda_\mathrm{FF})^2$ when comparing with
\d0. Here, $s$ is the LEP c.m. energy.  This is {\it not} done in
Fig.~\ref{fig:tgc}, since the effect only amounts to a few percent
increase. As can be seen, the \d0\ results are comparable to those
from the LEP experiments, and because of the different production mechanisms and
effects of anomalous couplings, the results are complementary.

\section{\boldmath CDF observation of $Z \to b \bar b$}

The $Z \to b \bar b$ decay channel, copious at LEP and SLC, is
challenging to observe at the Tevatron due to the large QCD dijet
background. However, it is an important benchmark signal for future
physics at the Tevatron -- it can be used to improve the jet
resolution in Higgs searches involving the $H \to b \bar b$ decay, and
it will be an important calibration tool for the top quark mass
measurement.

\begin{figure}
 \epsfysize = 7.5cm
    \centerline{\epsffile{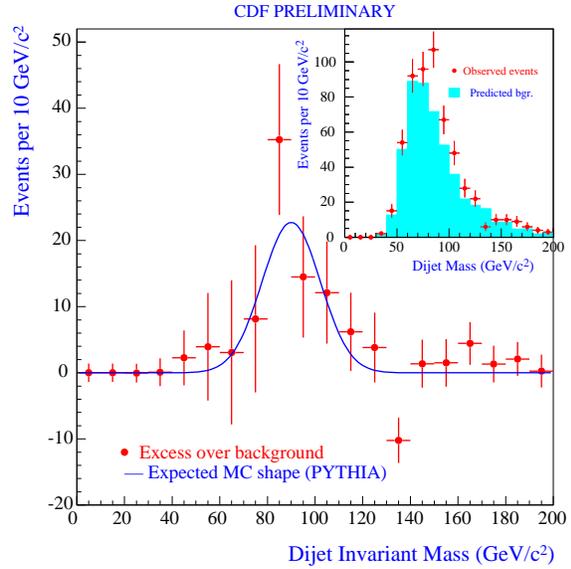}}
\caption{CDF $Z \to b \bar b$ invariant mass plot. The figure shows the dijet 
invariant mass for events containing a central muon and two SVX-tagged jets.}
\label{fig:zbb}
\end{figure}

In the CDF analysis, semi-leptonic decays of the $b$-quark are
utilized by selecting events from a sample collected with a central
muon trigger. Offline, a good muon candidate with $p_T$~$>$~$7.5$~GeV is
required, together with two jets containing charged tracks forming a
well-identified, displaced secondary vertex in the Silicon Vertex
Detector.  Topological cuts are applied to reject QCD background,
based on the amount and topology of the radiation surrounding the two
leading jets.  Figure~\ref{fig:zbb} shows the
background-subtracted dijet invariant mass distribution.  The peak and
width of the distribution are as expected from Monte Carlo $Z \to b
\bar b$ signal. Accounting for the systematic uncertainty on the
background prediction (4\%), the significance of the excess over
background is 3.2 standard deviations.

\end{document}